\documentclass[twocolumn,prl,aps,superscriptaddress,longbibliography]{revtex4-1}

\usepackage{graphicx}
\usepackage{amsmath}
\usepackage{amssymb}
\usepackage{marvosym}
\usepackage{txfonts}
\usepackage{bm}
\usepackage{xcolor}
\usepackage{color}

\newcommand{\lsco}{La$_{2-x}$Sr$_x$CuO$_4$}
\newcommand{\tbcod}{Tl$_2$Ba$_2$CuO$_{6+\delta}$}
\newcommand{\tbco}[1]{Tl$_2$Ba$_2$CuO$_{#1}$}
\newcommand{\ybco}[1]{YBa$_2$Cu$_3$O$_{#1}$}
\newcommand{\tc}{$T_c$}
\newcommand{\dwave}{$d$-wave}

\definecolor{TlBlue}{RGB}{31, 119, 180}
\def\TlColor{\color{TlBlue}}

\definecolor{LSCORed}{RGB}{214,39,40}
\def\LSCOColor{\color{LSCORed}}

\begin{document}

\title{From Mott to not: phenomenology of overdoped cuprates}

\author{N.~R.~Lee-Hone}
\affiliation{Department of Physics, Simon Fraser University, Burnaby, BC, V5A~1S6, Canada}
\author{H. U. \"Ozdemir}
\affiliation{Department of Physics, Simon Fraser University, Burnaby, BC, V5A~1S6, Canada}
\author{V. Mishra}
\affiliation{Computer Science and Mathematics Division, Oak Ridge National Laboratory, Oak Ridge, TN 37831, USA}
\author{D.~M.~Broun}
\affiliation{Department of Physics, Simon Fraser University, Burnaby, BC, V5A~1S6, Canada}
\affiliation{Canadian Institute for Advanced Research, Toronto, ON, MG5 1Z8, Canada}

\author{P.~J. Hirschfeld}
\affiliation{Department of Physics, University of Florida, Gainesville FL 32611}

\begin{abstract}
Recently, we have argued that experimental data on superfluid density and terahertz conductivity of overdoped LSCO are compatible with a Landau Fermi liquid/Bardeen-Cooper-Schrieffer description of these samples, provided dopants are treated within ``dirty $d$-wave theory" as weak scatterers.  Here we test these ideas by comparing to specific heat and thermal conductivity data on LSCO, showing that the theory works extremely well across the overdoped region for similar disorder parameters.   We then study the same properties in another overdoped cuprate, Tl-2201, thought to be quite ``clean" since it exhibits quantum oscillations, low residual resistivities and small superconducting state Sommerfeld coefficients.  Our results are consistent with the \mbox{Tl-2201} system being $\approx 3$ times cleaner due in part to the dopant atoms' being located further from the CuO$_2$ plane.  We conclude that cuprates can be described semiquantitatively in the overdoped regime by ``dirty $d$-wave" theory, subject to significant Fermi liquid renormalizations,  without introducing physics beyond the Landau-BCS paradigm. 
\end{abstract}

\maketitle{}

Of the various phases observed in the hole-doped cuprate  phase diagram \cite{Keimeretal:2014}, the one which seems most conventional in many ways is the $d$-wave superconductor between hole concentrations of roughly $p_{c1}\simeq 5\%$ and  $p_{c2}\simeq 30\%$.  Over much of this range, from onset to slightly past optimal doping for superconductivity at around 16\%, superconductivity condenses out of a state that is poorly understood, characterized by a pseudogap in the one-particle spectrum \cite{Timusk:1999p422}, unusual transport properties \cite{Gurvitch:1987ck,Thomas:1988gj,Varma:1989fr}, and  several other coexisting symmetry breaking orders \cite{TRANQUADA:1995p554,Lake:2001p655,Kivelson:2003p598,Sachdev:2003p428}.  On the other hand, if one is able to dope past around 20\%, into the so-called overdoped region, most of these effects seem to disappear and one can imagine solving the much simpler problem of $d$-wave superconductivity condensing out of a Fermi liquid.

It is worthwhile recalling that shortly after the discovery of cuprate superconductivity, the existence of $d$-wave superconductivity did not seem likely at all, as it was quite widely believed that disorder would destroy it due to the pairbreaking character of impurities in higher angular momentum pair states \cite{Balian:1963kc}.   Doping takes place for the most part via chemical substitution or oxygen removal in various layers away from the CuO$_2$ planes, and in the process almost always introduces disorder.  In the early 90s, studies showed theoretically that the presence of small amounts of  disorder in cuprates was both compatible with the existence of the $d$-wave state --- particularly because dopants  created rather weak scattering potentials as sensed by electrons moving in the plane --- and also rather important for understanding the observed properties \cite{Arberg:1993cu,Hirschfeld:1993cka,HIRSCHFELD:1993p567,HIRSCHFELD:1994p570}.    This theoretical approach, similar to the Abrikosov--Gor'kov theory of disordered conventional superconductors \cite{AAAbrikosov:1960}, now goes under the name of ``dirty $d$-wave'' theory.     Still, most discussions of the overdoped phases have ignored these effects, focusing instead on possible intrinsic physics despite the fact that  overdoping by chemical substitution necessarily introduces higher  concentrations of impurities.

A key question in this context is the mechanism responsible for the reduction in $T_c$ on overdoping, and the disappearance of superconductivity at $p_{c2}$.  For the most part, this has been attributed to intrinsic effects: it has been assumed that the strength of the electronic correlations in the CuO$_2$ plane responsible for pairing weakens as one goes to higher doping \cite{Scalapino_Maier2017}, due both to the enhanced screening of the local Coulomb interaction, and the easing of the Mott constraint on hole kinematics.     However, a few authors have discussed the role of disorder  in suppressing $T_c$ \cite{Balatsky:2006p607}, and this scenario was particularly emphasized by Rullier-Albenque et al.\ \cite{Rullier-Albenque2008}, who pressed the analogy between systematic irradiation  disorder, which suppresses $p_{c2}$ in YBCO, and dopant disorder.  

 Another longstanding puzzle is the apparent contradiction between the   ``universal" physics of the CuO$_2$ planes \cite{AndersonScience1987,Keimeretal:2014} and the wide variation of maximum $T_c$'s  across  cuprate
families.   An obvious and strong correlation is the number of planes per unit cell, but even within the single-layer cuprates the maximum $T_c$ varies from roughly 10~K for Bi-2201 to 40~K for LSCO to 90~K for Hg-1201 and Tl-2201.  Some authors have provided explanations in terms of band structure differences driven by apical oxygen states \cite{Pavarini2001}  or admixtures of additional Cu $d$ orbitals at the Fermi surface of different materials \cite{Sakakibara2010}.  However it is also true that various single-layer cuprates are doped in rather dissimilar ways.  Fujita et al.\ \cite{Fujita2005} proposed that the strength of the potential scattering introduced by the dopants in individual cuprates might account for most of the $T_c$ variation.  

In this paper, we study the role  disorder plays in these phenomena with the tools of dirty $d$-wave and Fermi liquid theory. We extend our previous work on the electrodynamic response of LSCO, where we found that calculations carried out using an accurate model of the Fermi surface, based on ARPES tight-binding dispersions, combined with realistic estimates of the disorder level, gave good agreement with both superfluid density \cite{Lee-Hone:2017} and terahertz conductivity \cite{Lee-Hone:2018}. Significant effective band renormalizations were required (see Supplemental Material), but these were consistent with those reported by other experiments \cite{Padilla:2005ir}.  The purpose of our current paper is to more broadly test the dirty $d$-wave model, which we do in two ways: by showing that it provides a good account of other experimental quantities (residual specific heat, thermal conductivity); and by extending the comparison to the Tl-2201 system, another single-layer cuprate that can be tuned throughout the overdoped regime.  The latter step provides a particularly stringent test of the model, as Tl-2201 is thought to be a very clean system, due to its manifestation of quantum oscillations \cite{Vignolle:2008p1694,Bangura:2010p1675,Rourke:2010bl} and the observation of clean-limit behavior in other properties \cite{MACKENZIE:1993p197,Proust:P2lqZi4f,Hussey:2003}.  We show that despite being indeed significantly cleaner than LSCO, overdoped Tl-2201 is describable within the same dirty $d$-wave framework, and is sufficiently dirty to display the same unusual ``non-BCS" like proportionality of superfluid density to $T_c$, as recently measured in overdoped LSCO \cite{Bozovic:2016ei}. Finally, we discuss the implications of our theoretical description of the overdoped state. 
\vskip .2cm

{\it Dirty $d$-wave theory.}
\label{theory}
The history and structure of the dirty $d$-wave theory have been reviewed recently in Ref.~\onlinecite{Lee-Hone:2018}. It is based on the single-particle Green's function 
\begin{eqnarray}
G(\mathbf{k},i\omega_n)=-\frac{i\tilde\omega_n \tau_0 + \Delta_{\bf k} \tau_1 + \xi_{\bf k} \tau_3} {\tilde\omega^2_n + \Delta_{\bf k}^2 + \xi_{\bf k}^2}\;,
\end{eqnarray}
where $\Delta_\mathbf{k}$ is the $d$-wave superconducting gap at \mbox{wave-vector $\mathbf{k}$}, $\xi_{\bf k}$ is the single-particle energy, $\tau_i$ are the Pauli matrices in Nambu space, and $\tilde \omega_n$ is a renormalized Matsubara frequency that, in the self-consistent $t$-matrix approximation (SCTMA) \cite{Hirschfeld:1986ii, Schmitt-Rink:1986}, obeys 
\begin{align}
\tilde \omega_n & =\omega_n + i\Sigma_\mathrm{imp}(\tilde\omega_n)=\omega_n + \pi \Gamma \frac{\langle N_\mathbf{k}(\tilde \omega_n) \rangle_\mathrm{FS}}{c^2 + \langle N_\mathbf{k}(\tilde \omega_n) \rangle_\mathrm{FS}^2}\;.
\label{tmatrix}
\end{align}
Impurity scattering is assumed to be characterized by parameters $(\Gamma,c)$, where \mbox{$c$ is the cotangent} of the scattering phase shift and $\Gamma$ is a scattering rate parameter proportional to the concentration of impurities.  The corresponding normal-state scattering rate is $\Gamma_N = \pi\Gamma/(1 + c^2)$.
$N_\mathbf{k}(\tilde \omega_n) = {\tilde \omega_n}/\big(\tilde \omega_n^2 + \Delta_\mathbf{k}^2\big)^{1/2}$ 
is the integrated diagonal Green's function, and 
 $\langle ... \rangle_\mathrm{FS}$ is a Fermi surface average, defined in Supplemental Material by Eqs.~S1 and S2.
The critical temperature and the temperature dependence of the order parameter are obtained by solving the gap equation
\begin{equation}
\Delta_\mathbf{k} =2 \pi T N_0 \sum_{\omega_n > 0}^{\omega_0} \left\langle V_{\mathbf{k},\mathbf{k}^\prime} \frac{\Delta_{\mathbf{k}^\prime}}{\sqrt{\tilde \omega_n^2 + \Delta_{\mathbf{k}^\prime}^2}}\right\rangle_\mathrm{FS}\;,
\label{gap_equation}
\end{equation}
where
$N_0$ is the total density of states, $V_{\mathbf{k},\mathbf{k^\prime}}$ is the  pairing interaction, and $\omega_0$ is a high energy cutoff.

 \begin{figure}[t]		
 \includegraphics[width=\columnwidth]{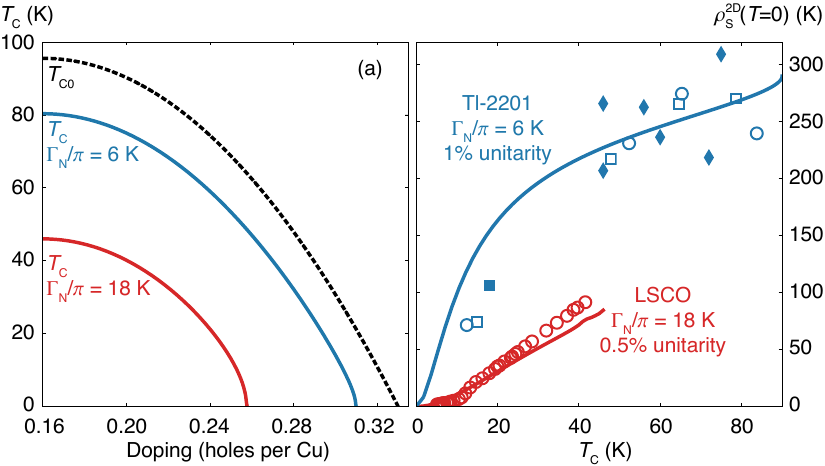}
 \caption{Predictions of dirty $d$-wave theory: (a) For a single parabolic doping dependence of the underlying $T_{c0}(p)$, different choices of $\Gamma_N$ result in superconducting domes $T_c(p)$ reminiscent of Tl-2201 and LSCO.  (b) Starting from accurate parameterizations of the Fermi surfaces, the same theory captures the strong correlation between $\rho_s$ and $T_c$ observed in experiment. LSCO data: {\LSCOColor \CircPipe} MBE thin-film mutual inductance \cite{Bozovic:2016ei}.  Tl-2201 data: $\TlColor\blacksquare$ single crystal microwave \cite{Deepwell:2013uu}; \mbox{$\TlColor\blacklozenge$ single crystal $\mu$SR \cite{Brewer2015};} {\TlColor\CircPipe}, {\TlColor\SquarePipe} polycrystalline $\mu$SR \cite{Niedermayer1993,Uemura1993}.}\label{fig1}
\end{figure}

{\it Comparison of LSCO and Tl-2201.}  In Refs.~\onlinecite{Bozovic:2016ei} and \onlinecite{Mahmood:2017}, arguments were given why disorder could not be the cause of the unusual superconducting behavior observed in overdoped LSCO films.  Principal among these were the linearity of the penetration depth measured to relatively low temperatures, but this turns out to be a feature of $d$-wave superconductors in the presence of weak scatterers, as pointed out in Refs.~\onlinecite{Kogan:2013kf} and \onlinecite{Lee-Hone:2017} and plotted in Fig.~S1. 
As discussed in the Supplemental Material, the Born limit is adequate to discuss impurity potentials $V_\mathrm{imp}\lesssim 0.1$ eV (i.e., $c \gtrsim 2$), which we believe includes out-of-plane chemical substituents and interstitials in cuprates. 
Furthermore, a disorder-based explanation of the properties of overdoped LSCO is perhaps not unexpected,  since many other characteristics of this  system are consistent with dirty limit behavior, e.g., the failure to observe quantum oscillations. 

 \begin{figure*}[t]			\includegraphics[width=0.9\textwidth]{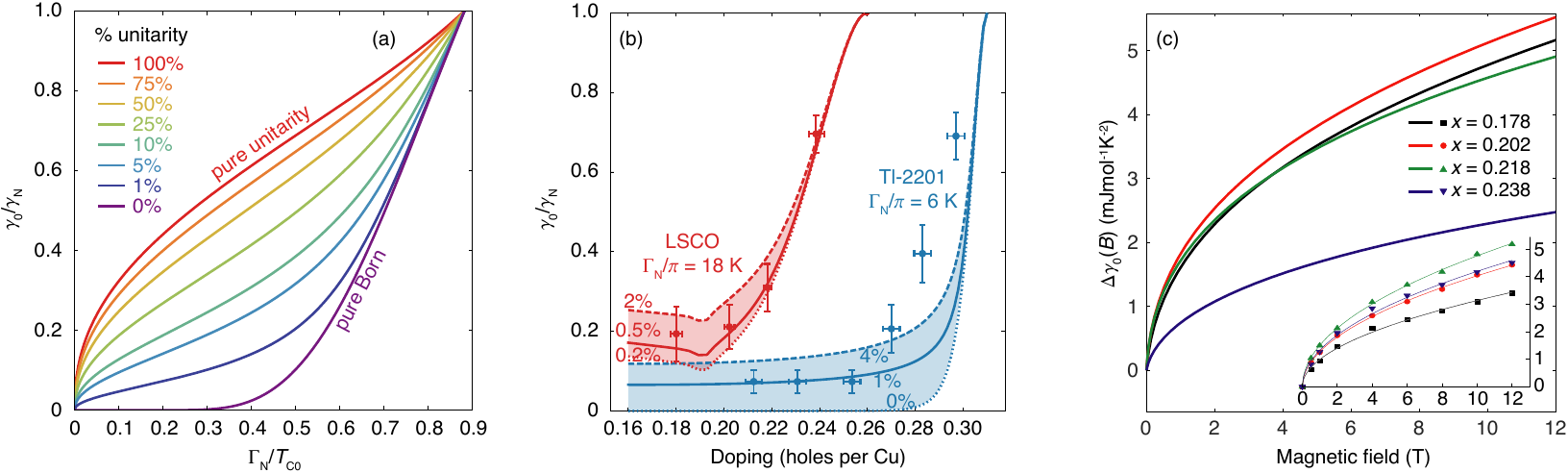}
			\caption{Specific heat in the dirty $d$-wave theory. (a) The residual Sommerfeld coefficient, $\gamma_0$, is a particularly sensitive probe of the degree of unitarity scattering in a $d$-wave superconductor. Curves are parameterized by the percentage contribution unitarity scattering makes to the normal-state scattering rate $\Gamma_N$. (b) Comparison of $d$-wave theory with zero-field heat capacity data for LSCO \cite{Wang2007} and Tl-2201 \cite{Loram1994}. (For Tl-2201 $\gamma(T)$ has been extrapolated to $T = 0$ using an equal entropy construction --- see Supplemental Material).  Errors bars denote experimental uncertainties.  For fixed total normal-state scattering rate, $\Gamma_N$, dirty $d$-wave results are plotted for varying percentages of unitarity scattering, as indicated.  (c) Calculated field dependence of residual Sommerfeld coefficient in the Doppler-shift approximation, for a square vortex lattice in LSCO \cite{Chang2012}, at four different dopings, for $\Gamma_N/\pi = 18$~K and 0.5\% unitarity scattering.  Inset: corresponding experimental data from Ref.~\onlinecite{Wang2007}.}
\label{fig2}
\end{figure*}

As mentioned above, by many measures Tl-2201 is a much cleaner cuprate system.   However, it has been established for some time that the superfluid density correlates strongly with $T_c$, as expected only in a dirty BCS superconductor \cite{Uemura1993,Niedermayer1993,Deepwell:2013uu,Brewer2015}. 
It therefore behooves us to consider the Tl-2201 system  as an important test of dirty $d$-wave theory and its ability to explain the nonuniversal aspects of the overdoped cuprate families.   
 
First, we ask, how clean is Tl-2201 really?  Estimates of the Dingle temperature in a $T_c = 27$~K ($p=0.27$) sample give a single-particle electronic mean free path in the normal state of about 360~\AA\ \cite{Rourke:2010bl}, compared to a transport mean free path of 620~\AA\ from microwave measurements \cite{Deepwell:2013uu}.    This suggests a modest amount of forward scattering character of the out-of plane defects in the Tl system, which we ignore for the moment but will discuss below.  By contrast, the transport mean free path deduced from the Mahmood et al.\ \cite{Mahmood:2017} terahertz measurements on LSCO is of order 100~\AA, 5-6 times smaller than in Tl-2201.  This factor is roughly consistent with comparisons of scattering rates in the two systems by Bangura et al.\ \cite{Bangura:2010p1675}, and also qualitatively consistent with the proposal of Fujita et al.\ \cite{Fujita2005} that those materials with A-site dopant disorder suffer from more pairbreaking than those where dopants reside an additional layer distant from the CuO$_2$ plane. Experimentally, it is known that the dominant source of cation disorder in Tl-2201 is an approximately 10\% excess of Cu that substitutes for Tl \cite{Shimakawa:1993jm}.  The TlO layers in which this disorder resides are $\approx 3$ times further from the CuO$_2$ planes than the LaO layers in LSCO in which the dopant Sr atoms reside.
 
 We therefore examine the predictions of the dirty \mbox{$d$-wave} theory for weak-to-intermediate dopant-type disorder, in a system several times cleaner than LSCO.  
 In Fig.~(1a) we show how two very different $T_c(p)$ relations, corresponding approximately to those of \mbox{Tl-2201} and LSCO, can emerge from a single clean-limit reference system (single $T_{c0}(p)$ curve), using the Abrikosov--Gor'kov $T_c$ suppression, with $\Gamma_N/\pi = 18$~K for the LSCO-like curve, as in earlier work \cite{Lee-Hone:2017,Lee-Hone:2018} and $\Gamma_N/\pi = 6$~K for Tl-2201.  Note that a larger critical doping naturally emerges for Tl-2201 ($p_{c2} = 0.31$) than for LSCO ($p_{c2} = 0.26$), in accord with experiment \cite{Vignolle:2008p1694,Bangura:2010p1675,Rourke:2010bl}.
 
 In Fig.~1(b), we plot the two-dimensional superfluid density
\begin{equation}
\rho_s^\mathrm{2D}(T) = \frac{\hbar^2 d}{4} 2 \pi T N_0\sum_{\omega_n > 0}\left\langle {v}_{F,x}^2\frac{\Delta_\mathbf{k}^2}{(\tilde \omega_n^2 + \Delta_\mathbf{k}^2 )^\frac{3}{2}} \right\rangle_{\mathrm{FS}}\;,
\label{superfluiddensity}
\end{equation}
 as calculated in Ref.~\onlinecite{Lee-Hone:2017} using the LSCO disorder parameters, together with the same calculation for parameters appropriate for Tl-2201. The tight-binding parameterizations of the Fermi surfaces  \cite{Yoshida:2006hw,Plate:2005} are discussed in Supplemental Material.  We note in particular that the Tl-2201 parameterization is based directly on low energy ARPES and therefore requires no additional renormalization of the dispersion.  It is seen that the dirty $d$-wave model describes both the cleaner Tl-2201 system and the dirty LSCO system quite well, nevertheless capturing the observed ``non-BCS" scaling of $\rho_s$ with $T_c$ in the Tl-system despite this behavior being associated with significant disorder in BCS theory.  

Since cuprates are famous for displaying behavior that deviates strongly from that of Fermi liquids, making a case for a conventional description in the overdoped region of the phase diagram requires further testing and comparison with additional data.  In Fig.~2, we display the results of an evaluation of the superconducting state specific heat $C(T)=TdS/dT$, obtained from the Bogoliubov quasiparticle entropy \cite{HIRSCHFELD:1988p611}
\begin{equation}
S= - k_B \int d\omega N(\omega)\left[f \ln f - (1- f) \ln(1 - f)\right]\;,
\label{entropy}
\end{equation}
where $f = f(\omega,T)$ is the Fermi function. In Fig.~\ref{fig2}(a), we  illustrate how  the  residual Sommerfeld coefficient \mbox{$\gamma_0 = \lim_{T\rightarrow 0}C(T)/T$}, reflecting the  density of states $N(0)$, depends on scattering phase shift and is dominated mostly by the strong scatterers.   This comparison is thus the most sensitive way to pin down the magnitude of the near-resonant disorder scattering in the CuO$_2$ planes.
In Fig.~\ref{fig2}(b) we present comparisons of the theory with existing data on doping-dependent Sommerfeld coefficients of LSCO \cite{Wang2007} and Tl-2201, where the latter have been obtained using entropy-conserving fits to the data in Ref.~\onlinecite{Loram1994}, as described in Supplemental Material. It is seen that the appropriate values of the unitarity limit scattering rate parameter are somewhat smaller than used in our earlier comparisons: this does not change any of the fits or conclusions of those studies, as the electromagnetic response is far less sensitive to strong scattering than the heat capacity.  Overall, the dirty $d$-wave model fits the doping dependence of the experimental data on both systems extremely well. 

The effect of a magnetic field in the superconducting state can be approximated by including the Doppler shift of the quasiparticle energy in the semiclassical approximation to the density of states \cite{VOLOVIK:1993p201,Kuebert1998,Ioffe:2002p631},
\begin{equation}
\label{dos with field}
N( \omega,H) =
N_0 \:\mathrm{Im} \left\langle\left\langle \frac{\tilde \omega\left(\omega - \tfrac{1}{2}\mathbf{p}_s\cdot \mathbf{v}_\mathbf{k}\right)}
 {\sqrt{\tilde \omega^{\,2}\!\left(\omega - \tfrac{1}{2}\mathbf{p}_s\cdot \mathbf{v}_\mathbf{k}\right) - \Delta_\mathbf{k}^2 }}\right\rangle_\mathrm{\!\!FS}\right\rangle_{\!\!R}
\;.
\end{equation}
Here $\mathbf{p}_s$ is the local gauge-invariant superfluid momentum at point $R$ and $\langle\rangle_R$ is a spatial average over the vortex lattice unit cell, as described in detail in Supplemental Material. In Fig.~\ref{fig2}(c),  we show the calculated field dependence of the superconducting state Sommerfeld coefficient for a range of LSCO dopings, with the same impurity parameters used above.  The most striking aspect is the persistence of the approximate $\gamma\sim \sqrt{H}$ behavior in  relatively dirty samples.  In addition, the overall magnitude of the field variation is quite consistent with the experimental data from Ref.~\onlinecite{Wang2007}. In the original paper, by focussing on the strong scattering regime of the dirty $d$-wave theory, the experimentalists were unable to reconcile the relatively large $\gamma_0$ with the observation of apparently clean-limit $\sqrt{H}$ behaviour,
and concluded that the system must be undergoing real-space phase separation.  In fact, we see here that this apparent  inconsistency is explained naturally when one accounts for the presence of mostly weak scatterers and a small concentration at the unitarity limit. 

\begin{figure}[t]	
 \includegraphics[width=0.65\columnwidth]{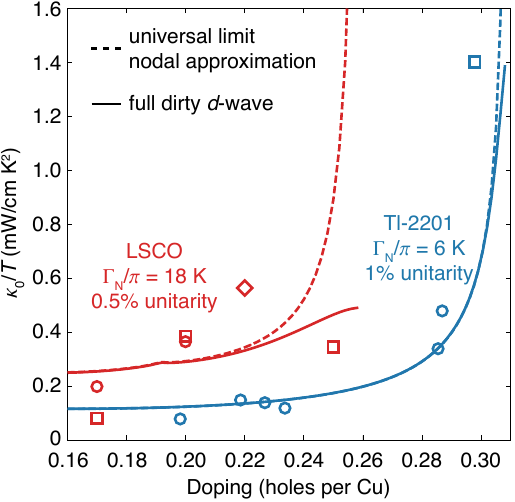}	\caption{Thermal conductivity data on LSCO 
({\LSCOColor \CircPipe} \cite{Takeya:2002hh}, {\LSCOColor \SquarePipe} \cite{Sutherland:2003p753}) and \mbox{Tl-2201} ({\TlColor \CircPipe} \cite{Hawthorn:2007hq}, {\TlColor \SquarePipe} \cite{Proust:P2lqZi4f}) 
 compared to dirty $d$-wave theory with appropriate parameters, as discussed in the text. Note the breakdown of the universal-limit nodal approximation on the approach to $p_{c2}$ in LSCO, but not in Tl-2201.}
\label{fig:thermalcond}
\end{figure}

Next, we discuss the thermal conductivity in the superconducting state, a sensitive probe of the lowest energy mobile $d$-wave quasiparticles, with
\begin{equation}
\!\,\frac{\kappa}{T} \!=\!\frac{N_0}{2}
\!\!\int_{0}^{\infty} \!\!\!\!\!\!d\omega \frac{\omega^2}{T^2} \frac{\partial f}{\partial \omega} \left \langle
\frac{v_{F,x}^{2}}{{\rm
Re}\!\sqrt{\tilde{\Delta}_{\bf k}^{2}-\tilde{\omega}^{2}}}
\left[\frac{|\tilde{\Delta}_{\bf k}|^2 - |\tilde{\omega}|^2}{|
\tilde{\Delta}_{\bf k}^{2}-\tilde{\omega}^{2}|}-1\right]\right
\rangle_\mathrm{\!\!\!FS}\!\!,
\label{eq:kappa1} 
\end{equation}
    which reduces at low $T$ and $\Gamma_N$ to the universal limit \mbox{$\kappa_{00}/T\equiv k_B^2 v_F /(3 \hbar d v_\Delta)
$}, where the gap velocity $v_\Delta$ includes the effects of disorder via self-consistent solution of the BCS gap equation, Eq.~\ref{gap_equation}.  In Fig.~\ref{fig:thermalcond}, we compare evaluations of Eq.~\ref{eq:kappa1} for the same LSCO and Tl-2201 disorder parameters and band structures to the available experimental data.   Note that the full theory distinguishes between the rapid rise of $\kappa_0/T$ as $p \to p_{c2}$ in Tl-2201 and the weak doping dependence in LSCO.  The consistency is particularly impressive given the lack of adjustable parameters.

 \begin{figure}[t]			
 \includegraphics[width=\columnwidth]{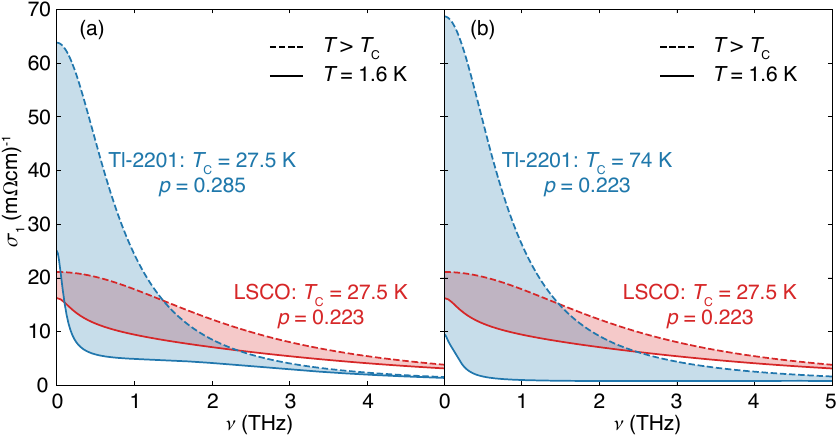}
\caption{Comparison of the optical conductivity of overdoped LSCO and Tl-2201, calculated in the dirty $d$-wave model for (a) materials with the same $T_c$ but different doping level; and (b) materials with the same doping level but different $T_c$. Shaded regions denote the spectral weight that condenses to form the low-temperature superfluid. Impurity parameters are the same as earlier in the paper:  $\Gamma_N/\pi = 6$~K for Tl-2201, with 1\% unitarity scattering; and $\Gamma_N/\pi = 18$~K for LSCO, with 0.5\% unitarity scattering.}
\label{fig4}
\end{figure}

 Finally, in Fig.~\ref{fig4} we calculate $\sigma_1(\Omega)$ in LSCO, (updated from Ref.~\onlinecite{Lee-Hone:2018} to incorporate the new, lower degree of unitarity scattering), and compare to expected results for Tl-2201.   The conductivity at frequency $\Omega$ is given by
\begin{equation}
 \sigma_1 \! =\! \frac{N_0 e^2}{2 \Omega}\!\! \int_{-\infty}^{\infty} \!\!\!\!\!d\omega \Big( f(\omega\!+\!\Omega)\!-\!f(\omega)\Big)
 \left\langle\!v_{F,x}^2\mathrm{Re}\Big\{ A_{++}\!\!-\!\!A_{+-}\Big\}\right\rangle_\mathrm{FS},
\label{eq:conductivity}
\end{equation}
where $A_{+\pm} = \left(\Delta_{\bf k}^2+\tilde\omega_+ \tilde\omega'_\pm + Q_+ Q' _\pm \right)/\Big( Q_+ Q '_\pm \left(  Q_+ + Q '_\pm\right)\Big)$, 
$\omega_\pm = \omega \pm i \eta$, $\omega_\pm'=\omega_\pm(\omega+\Omega)$,
$Q_\pm = (\Delta_{\bf k}^2-\tilde \omega^2_\pm)^{1/2}$, and $Q_\pm'=Q_\pm(\omega+\Omega)$.  Vertex corrections do not appear because we have assumed zero-range scatterers. To our knowledge, no THz conductivity data are yet available on the Tl-2201 system, so this may be regarded as a prediction of the theory.  As can be seen, in Tl-2201 the $\Omega\rightarrow 0$ conductivity in the normal state is significantly higher, and the degree of residual, uncondensed spectral weight as \mbox{$T \to 0$} significantly lower, both consistent with the lower level of disorder scattering. In addition, a narrow low-frequency component is clearly visible in Tl-2201 at low temperatures, but is not particularly prominent in LSCO.  In neither system does the gap edge correspond to any observable feature in the conductivity spectrum, although it may indirectly appear via a ``4$\Delta$" feature if spin fluctuations are included in the cleaner Tl-2201 case \cite{Hirschfeld:1996}.  Inclusion of forward-scattering corrections, which could be important in the Tl-2201 system, may influence these results quantitatively.

{\it Conclusions.}  
General statements about any part of the ``cuprate phase diagram" are a priori dangerous because cuprates  consist of differing numbers of CuO$_2$ planes per unit cell, as well as differing charge reservoir layers.  The expectation, based on analysis of simple models, that correlations should weaken and that Fermi liquid properties should become clearly observable as one overdopes, has proven difficult to verify.  While normal state quasiparticle features are now routinely observed by ARPES in some systems at both nodal and antinodal points, some classic manifestations of normal state Fermi liquid behavior like $T^2$ resistivity do not appear to be realized over significant temperature ranges \cite{Taillefer_AnnRev:2010}.  This may be due to singular self-energy effects \cite{MiyakeVarma:2018}, but also quite likely to the presence of inelastic scattering from nearly condensed fluctuating order.  In any case, in the superconducting state the quasiparticle scattering rate generally collapses, leading to well-defined Bogoliubov quasiparticles, so the possibility to observe the underlying Fermi liquid is enhanced.   Our calculations  show that the Landau--BCS paradigm  provides a very adequate description of the low energy phenomenology of the overdoped cuprates, provided that the starting point is an accurate parameterization of the electronic dispersion, and that the occasionally nonintuitive effects of disorder are accounted for.  This, in turn, suggests that no exotic new physics should be required to understand the occurence of superconductivity starting from a weak-coupling approach on the overdoped side.

{\it Acknowledgements.} We are grateful for useful discussions with L.~Benfatto, A.~Damascelli, T.~P.~Devereaux, J.~S.~Dodge, D.~G.~Hawthorn, M.~P.~Kennett, S.~A.~Kivelson, C.~P\'epin and J.~E.~Sonier.   We gratefully acknowledge financial support from the Natural Science and Engineering Research Council of Canada, the Canadian Institute for Advanced Research, and the Canadian Foundation for Innovation.  P.~J.~H.\ was supported by the Department of Energy under \mbox{Grant No.~DE-FG02-05ER46236}.
\phantom{\cite{MOLER:1994p209,Clem:1975gr,Rourke2011,BPT}}

\end{document}